\documentclass[aps,prb,twocolumn,showpacs,showkeys]{revtex4}
\usepackage{graphicx}

\begin{document}

\title{
Isotopic shift in angle-resolved photoemission spectra of
Bi$_2$Sr$_2$CaCu$_2$O$_8$ due to quadratic electron-phonon coupling
}

\author{Kai Ji}
\affiliation{
Solid State Theory Division, Institute of Materials Structure Science, KEK,
Graduate University for Advanced Studies, Oho 1-1, Tsukuba, Ibaraki 305-0801,
and CREST, JST, Japan
}

\author{Keiichiro Nasu}
\affiliation{
Solid State Theory Division, Institute of Materials Structure Science, KEK,
Graduate University for Advanced Studies, Oho 1-1, Tsukuba, Ibaraki 305-0801,
and CREST, JST, Japan
}

\date{\today}

\begin{abstract}
In connection with the experiment on oxygen isotope effect of
Bi$_2$Sr$_2$CaCu$_2$O$_8$ with the angle-resolved photoemission spectroscopy
(ARPES), we theoretically study the isotope-induced band shift in ARPES
by the Hartree-Fork and quantum Monte Carlo methods.
We find that this band shift can be clarified based on a quadratically
coupled electron-phonon ($e$-ph) model.
The large ratio of band shift versus phonon energy change is connected
with the softening effect of phonon, and the positive-negative sign change
is due to the momentum dependence of the $e$-ph coupling.
\end{abstract}

\pacs{71.38.-k, 74.25.Kc, 74.72.-h, 79.60.-i}

\keywords{quadratic electron-phonon interaction, isotope effect,
angle-resolved photoemission spectra}

\maketitle

\section{Introduction}

The study of superconductivity has a long history.
For the conventional superconductors, the mechanism has been successfully
explained by the BCS theory\cite{bcs}, that the electron-phonon ($e$-ph)
interaction is at the central stage.
For the high-$T_{c}$ superconductors, however, the mechanisms is still
not clear with a number of basic questions need further investigation.
In order to clarify these problems, enduring efforts have been made for
two decades by the experimentalists and theorists.

In the experimental aspect, since the angle-resolved photoemission spectroscopy
(ARPES) directly measures the electronic occupied states in a crystal,
it has become one of the most important experimental techniques to study
the electronic properties of superconductors.
Nowadays, with the great improvement of energy resolution, ARPES is able
to observe the fine structures due to $e$-ph interaction\cite{la01} and the
tiny superconducting gap\cite{ki05} with an energy resolution less than 1 meV.
Recently, the oxygen isotope effect has been studied with ARPES on the
high-$T_c$ superconductor Bi$_2$Sr$_2$CaCu$_2$O$_8$ (Bi2212) by two
groups\cite{gw04,do07}.
Comparing their data, at least one common feature has become clear, the
spectra are shifted when $^{16}$O is substituted by its isotope $^{18}$O,
whose movement in this crystal is regarded as a lattice vibration, i.e.,
phonon.
This isotope induced band shift indicates that the $e$-ph interaction
has an effect on the electrons, hence providing direct evidence for the
interplay between electrons and phonons in this material.

Since the first report by Gweon {\it et al.}\cite{gw04}, the isotopic
band shift has become an controversial issue and gained considerable
interest\cite{ko04,ma05,fr05}, as the observed band shift is up to 40
meV, much larger than the maximum isotopic energy change of phonon ($\sim$
5 meV) according to the measured vibration energies of oxygen\cite{ma95}.
In addition to this large ration of band shift versus phonon energy change,
the presence of both positive and negative band shifts in the ARPES also
implies that the electrons and phonons are coupled in a complicated manner.
Very recently, Douglas {\it et al.}\cite{do07} repeat the experiment and
find the shift is only 2$\pm$3 meV, which is inconsistent with the large
shift found by Gweon {\it et al.}\cite{gw04}.

In this paper, we look into the $e$-ph interaction in Bi2212 and its
relation with ARPES by the Hartree-Fork and quantum Monte Carlo
(QMC)\cite{ji04} methods.
Our purpose here is to figure out whether the isotope-associated anomalies,
especially the large ratio of band shift versus phonon energy change,
are possible at all from a theoretical point of view.
We will show that the abovementioned band shift can be explained within
a scenario of phonon softening effect driven by the $e$-ph coupling, while
the positive-negative sign change is due to a momentum dependence of the
coupling.
The remainder of this paper is organized as follows.
In Sec. 2, we put forward the model Hamiltonian.
In Sec. 3, an investigation on this model with Hartree-Fork theory is
conducted.
In Sec. 4, we show the numerical results of Hartree-Fork and QMC methods,
and discuss the origin of band shift.
Our conclusion is presented in Sec. 5.

\section{Model}

In the theoretical aspect, to investigate the effect of lattice vibration
on the electrons, two kinds of $e$-ph interaction are usually modeled
for cuprates, i.e., diagonal and non-diagonal couplings\cite{is04}.
The former is a correlation between the phonons and intra-site electron
density\cite{ro04}, and the latter is between the phonons and inter-site
chemical bond\cite{sh02}.
In the CuO$_2$ plane of cuprates, as shown in Fig. 1, when an electron
is hopping from one copper atom to another, its movement is strongly
affected by the vibration of oxygen atoms between the initial and final
sites\cite{ku05}.
This kind of interaction thus turns out to be an off-diagonal type coupling,
which modulates the electronic band structure.
In this sense, the off-diagonal $e$-ph coupling is crucial to the
aforementioned isotope-induced band shift.

In this work, since we are primarily concerned with the effect of band
shift, we start from the following model Hamiltonian
($\hbar = 1$ and $k_B = 1$ throughout this paper):
\begin{eqnarray}
H &=& - {1 \over 2} \sum_{l, \sigma} \sum_{i, \delta_i} t(l, l + \delta_i)
  (a^{\dag}_{l, \sigma} a_{l + \delta_i, \sigma}
  + a^{\dag}_{l + \delta_i, \sigma} a_{l, \sigma})
  \nonumber\\
&& - \mu \sum_{l, \sigma} a^{\dag}_{l \sigma} a_{l \sigma}
  + {\omega_0 \over 2} \sum_{\langle l,l' \rangle} \left(-{1 \over \lambda}
  \frac{\partial^2}{\partial q^2_{ll'}} + q^2_{ll'} \right) ,
\end{eqnarray}
where $a^{\dag}_{l \sigma}$ ($a_{l \sigma}$) is the creation (annihilation)
operator of an electron with spin $\sigma$ at the Cu site $l$ on a square
lattice.
$t(l, l + \delta_i)$ is the electronic transfer energy between two Cu sites
$l$ and $l + \delta_i$, and $\delta_i$ enumerates the $i$th ($i$=1 or 2 in
this work) nearest neighboring Cu sites, which is accessible from site $l$.
$\mu$ is the chemical potential, defining the total electron number in this
system.
The oxygen phonon is assumed to be of the Einstein type with a frequency
$\omega_0$ and a mass $m$.
$\lambda$ ($\equiv 1 + \Delta m / m$)
is the mass change factor of phonon due to the isotope substitution.
In the third term, $q_{ll'}$ is the dimensionless coordinate operator
of the oxygen phonon located between the nearest-neighboring Cu sites
$l$ and $l'$, and the sum denoted by ${\langle l,l' \rangle}$ just means
a summation over all the phonon sites in the lattice.

\begin{figure}
\includegraphics{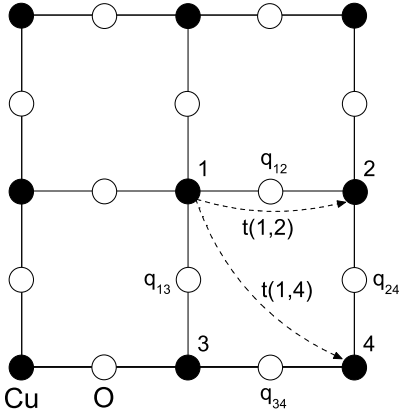}
\caption{
A Schematic plot of CuO$_2$ conduction plane in cuprates.
The copper atoms (black circles) form a simple square lattice, and the oxygen
atoms (white circles) are located between the nearest-neighboring Cu sites.
}
\end{figure}

In the conduction plane of CuO$_2$, the electronic hopping integral
$t (l, l')$ can be expanded to the second order terms with respect to
the phonon displacements.
For example, in Fig. 1, the nearest-neighbor hopping $t(1,2)$ and the
second-nearest-neighbor hopping $t(1,4)$ can be expanded as,
\begin{eqnarray}
t (1,2) &=& t_1 + s_1 q_{12}^2,
\\
t (1,4) &=& t_2 + s_2 (q_{12}^2 + q_{24}^2 + q_{13}^2 + q_{34}^2).
\end{eqnarray}
Here we note that the linear $e$-ph coupling does not appear in the
expansion owing to the lattice symmetry of present model.
Therefore, if we consider the hopping up to the second nearest neighbors,
Hamiltonian (1) can be rewritten as,
\begin{eqnarray}
H &=& - {1 \over 2} \sum_{l, \sigma} \sum_{\delta_1}
  ( t_1 + s_1 q_{l, l + \delta_1}^2 )
  ( a^{\dag}_{l, \sigma} a_{l + \delta_1, \sigma}
  + a^{\dag}_{l + \delta_1, \sigma} a_{l, \sigma} )
  \nonumber\\
&& - {1 \over 2} \sum_{l, \sigma} \sum_{\delta_2}
  \left[ t_2 + s_2 \sum_{\delta_1}'
  ( q_{l, l + \delta_1}^2 + q_{l + \delta_1, l + \delta_2}^2) \right]
  \nonumber\\
&& \mbox{\ \ \ \ \ } \times ( a^{\dag}_{l, \sigma} a_{l + \delta_2, \sigma}
  + a^{\dag}_{l + \delta_2, \sigma} a_{l, \sigma} )
  \nonumber\\
&& - \mu \sum_{l, \sigma} a^{\dag}_{l \sigma} a_{l \sigma}
  + {\omega_0 \over 2} \sum_{\langle l,l' \rangle} \left(-{1 \over \lambda}
  \frac{\partial^2}{\partial q^2_{ll'}} + q^2_{ll'} \right) ,
\end{eqnarray}
where $t_1$ ($t_2$) is the bare nearest- (second-nearest-) neighbor hopping
energy, and $s_1$ ($s_2$) is the off-diagonal $e$-ph coupling strength
due to the nearest- (second-nearest-) neighbor hopping effect.
The sum denoted by $\sum'_{\delta_1}$ in the second term of Eq. (4) is over
the common nearest neighbors shared by the Cu sites $l$ and $l + \delta_2$.
Here we want to stress that by using this model, we have assumed the
anti-ferromagnetic ordering is established as a background.
Accordingly, the double occupancy does not occur and the on-site Coulomb
repulsion between the electrons is omitted.
Mean while, the inter-site electron-electron interaction is partially
included in the screened values of $t$'s and $s$'s.

\section{Hartree-Fork treatment}

In order to have an insight in the effect of quadratic $e$-ph coupling,
we first study the model Hamiltonian (4) via a Hartree-Fork treatment.
By applying the Hartree-Fork approximation (HFA),
\begin{equation}
AB \rightarrow A \langle B \rangle + B \langle A \rangle
- \langle A \rangle \langle B \rangle ,
\end{equation}
we can separate the $e$-ph coupled term in Eq. (4) and get an effective
Hamiltonian as,
\begin{eqnarray}
H_{\rm HFA} &=& H_e + H_{\rm ph},
  \\
H_e &=& - {1 \over 2} \sum_{l, \sigma} \sum_{\delta_1}
  [ t_1 + s_1 \langle q^2 (\lambda) \rangle ]
  \nonumber\\
&& \mbox{\ \ \ \ \ } \times ( a^{\dag}_{l, \sigma} a_{l + \delta_1, \sigma}
  + a^{\dag}_{l + \delta_1, \sigma} a_{l, \sigma} )
  \nonumber\\
&& - {1 \over 2} \sum_{l, \sigma} \sum_{\delta_2}
  [ t_2 + 4 s_2 \langle q^2 (\lambda) \rangle ]
  \nonumber\\
&& \mbox{\ \ \ \ \ } \times ( a^{\dag}_{l, \sigma} a_{l + \delta_2, \sigma}
  + a^{\dag}_{l + \delta_2, \sigma} a_{l, \sigma} )
  \nonumber\\
&& - \mu \sum_{l, \sigma} a^{\dag}_{l\sigma} a_{l \sigma},
  \\
H_{\rm ph} &=& {\omega_0 \over 2} \sum_{\langle l,l' \rangle}
  \left[ -{1 \over \lambda}
  \frac{\partial^2}{\partial q^2_{ll'}} + (1 - \gamma) q^2_{ll'} \right],
\end{eqnarray}
where $\langle \cdots \rangle$ means the expectation value of an operator,
and
\begin{eqnarray}
\gamma &=& {1 \over 2 N \omega_0} \sum_{l, \sigma} \left(
  s_1 \sum_{\delta_1} \langle a^{\dag}_{l \sigma} a_{l + \delta_1, \sigma}
  + a^{\dag}_{l + \delta_1, \sigma} a_{l \sigma} \rangle \right .
  \nonumber\\
&& + \left .
  4 s_2 \sum_{\delta_2} \langle a^{\dag}_{l \sigma} a_{l + \delta_2, \sigma}
  + a^{\dag}_{l + \delta_2, \sigma} a_{l \sigma} \rangle \right).
\end{eqnarray}
As will be shown later, $\gamma$ acts as a renormalization factor for
the phonon energy.

According to the many-body theory, the spectral function can be obtained
from the single-electron Green's function\cite{ma90}.
After some algebra, we get the electronic spectral function from Eq. (7)
as,
\begin{eqnarray}
A_{\bf k} ( \lambda, \omega) = \delta [ \omega - E_{\bf k} (\lambda) ] ,
\end{eqnarray}
where $E_{\bf k} (\lambda)$ is the phonon-mediated tight binding energy
of an electron with momentum $\bf k$,
\begin{eqnarray}
E_{\bf k} (\lambda) &=&
  -2 [ t_1 + s_1 \langle q^2 (\lambda) \rangle ] (\cos k_x + \cos k_y)
  \nonumber \\
&&  -4 [ t_2 + 4 s_2 \langle q^2 (\lambda) \rangle ] \cos k_x \cos k_y
  - \mu .
\end{eqnarray}
Hence, the isotope induced band shift $\Delta E_{\bf k}$
[$\equiv E_{\bf k} (\lambda_0) - E_{\bf k} (\lambda)$]
has the following form,
\begin{eqnarray}
\Delta E_{\bf k} = -s_{\bf k} \Delta \langle q^2 \rangle,
\end{eqnarray}
where,
\begin{eqnarray}
s_{\bf k} = 2 s_1 (\cos k_x + \cos k_y) + 16 s_2 \cos k_x \cos k_y ,
\end{eqnarray}
is the momentum dependent coupling constant, and
\begin{eqnarray}
\Delta \langle q^2 \rangle \equiv \langle q^2 (\lambda_0) \rangle
  - \langle q^2 (\lambda) \rangle,
\end{eqnarray}
is the phonon spatial variation due to the
isotope substitution.
Here, one can clearly see that the band shift $\Delta E_{\bf k}$ is highly
anisotropic due to the momentum dependence of coupling $s_{\bf k}$.
This anisotropy can give rise to the above-mentioned sign reverse of the
isotope effect.
We shall return to this point in the next section.

Since the $e$-ph coupling is in a quadratic manner, the phonon wave function
retains a Gaussian form, and we assume it to be
\begin{eqnarray}
\phi_{ll'} = \sqrt{D \over \pi^{1/2}} \exp \left( - {D^2 q_{ll'}^2 \over 2}
  \right),
\end{eqnarray}
where $D$ is a coefficient to be determined by the variation method.
By using the Gaussian integral formula,
\begin{eqnarray}
& & \int_{- \infty}^{\infty} dq \exp (- \alpha q^2) = \sqrt{\pi \over \alpha},
  \\
& & \int_{- \infty}^{\infty} dq q^2 \exp (- \alpha q^2)
  = {1 \over 2} \sqrt{\pi \over \alpha^3},
\end{eqnarray}
we have
\begin{eqnarray}
& & \langle \phi_{ll'} | q_{ll'}^2 | \phi_{ll'} \rangle = {1 \over 2D^2},
  \\
& & \langle \phi_{ll'} | - {\partial^2 \over \partial q_{ll'}^2} | \phi_{ll'}
  \rangle = {D^2 \over 2},
\end{eqnarray}
and the phonon energy $E_{\rm ph}$
($\equiv \langle \phi_{ll'} | H_{\rm ph} | \phi_{ll'} \rangle$) is
\begin{eqnarray}
E_{\rm ph} = {N \omega_0 \over 2}
  \left( {D^2 \over \lambda} + {1 - \gamma \over D^2} \right).
\end{eqnarray}
Minimizing the phonon zero point energy of Eq. (20) with respect to $D$,
the coefficient $D$ is finally determined as
\begin{eqnarray}
D=\sqrt[4]{ (1 - \gamma) \lambda }.
\end{eqnarray}
Substituting Eq. (21) into Eq. (18), the phonon spatial variation is found to be
\begin{eqnarray}
\Delta \langle q^2 \rangle = {1 \over 2 \sqrt{1 - \gamma}} \left(
  {1 \over \sqrt{\lambda_0}} - {1 \over \sqrt{\lambda}} \right) .
\end{eqnarray}
Thus, within the HFA, we can self-consistently determine the isotopic band
shift in terms of Eqs. (9), (12), (22).

In terms of Eqs. (20) and (21), the phonon energy becomes
\begin{eqnarray}
E_{\rm ph} = {N \omega_s \over \sqrt{\lambda}} ,
\end{eqnarray}
where,
\begin{eqnarray}
\omega_s = \omega_0 \sqrt{1 - \gamma} ,
\end{eqnarray}
is the effective phonon energy renormalized by $\gamma$, which means the
phonon is softened when the $e$-ph coupling is switched on.
This phonon softening effect is not due to the photoemission or isotope
replacement, but the intrinsic $e$-ph coupling which is in the quadratic
manner.
Here we want to stress that the bare phonon energy $\omega_0$ is not an
experimental observable, meanwhile the effective one $\omega_s$ can be
detected by the Raman spectroscopy.
We now introduce the isotopic phonon energy change as
\begin{eqnarray}
\Delta \omega & \equiv & \omega_s (\lambda_0) - \omega_s (\lambda) \\
  & = & \omega_0 \sqrt{1 - \gamma} \left( {1 \over \sqrt{\lambda_0}}
  - {1 \over \sqrt{\lambda}} \right) .
\end{eqnarray}
From Eqs. (12), (22) and (26), we find the ratio
$\Delta E_{\bf k} / \Delta \omega$ can be represented as
\begin{eqnarray}
{\Delta E_{\bf k} \over \Delta \omega} =
  - {s_{\bf k} \over 2 \omega_0 (1 - \gamma)} ,
\end{eqnarray}
which is only subject to the $e$-ph coupling.
In this sense, $\Delta E_{\bf k} / \Delta \omega$ can be regarded as a
quantity indicative of the strength of $e$-ph coupling and the degree
of phonon softening in the $e$-ph system.

\section{Results and discussion}

As mentioned above, the most intriguing behavior of the isotope effect
is associated with the ratio $\Delta E_{\bf k} / \Delta \omega$.
We shall focus on this ratio in the present section.

In order to compare the HFA results with the exact calculation, we employ
the QMC method\cite{ji04} in this work.
Starting from Hamiltonian (4), we first calculate the spectral function
$A_{\bf k} (\lambda, \omega)$,
and then determine $E_{\bf k} (\lambda)$ by the moment analysis
of the spectral function as,
\begin{equation}
E_{\bf k} (\lambda) = \int^{\infty}_{- \infty}
  \omega A_{\bf k} ( \lambda, \omega ) d \omega .
\end{equation}
In QMC, the temperature is set at $T$=0.05.
Since this simulation suffers from the random error, in the numerical
calculation we impose a large isotopic mass enhancement to suppress the
fluctuation, i.e., from $\lambda_0$=1 to $\lambda$=2.
It should be noted here, although this mass variation is larger than the
$^{16}$O/$^{18}$O substitution, it actually has no effect on the ratio
$\Delta E_{\bf k} / \Delta \omega$, which is mass-independent, as shown
in Eq. (27).

\begin{figure}
\includegraphics{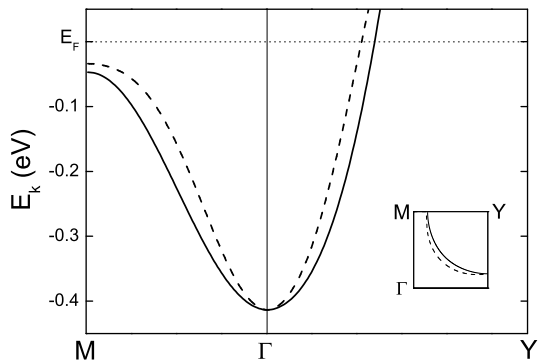}
\caption{
Energy dispersion of Bi2212 from the tight-binding fit of this work (solid
curve) and Ref. 16 (dashed curve).
(Parameter selection is described in the text.)
The inset shows a quadrant of the Brillouin zone with Fermi surface.
}
\end{figure}

In Ref. 16, five tight binding parameters are suggested for a best agreement
with the experimental band structure and Fermi surface of Bi2212.
While, in this work, since we are interested in the mechanism of band
shift, we restrict our discussion in a case with only nearest- and
second-nearest-hopping effects, and thus introduce two tight binding
parameters, i.e., the bare electronic transfer energies $t_1$ and $t_2$.
To reproduce the band structure and Fermi surface, we set $t_1$=0.22 eV
and make it the unit of energy for the numerical calculation.
In the unit of $t_1$, we assume $t_2$=-0.2, and the chemical potential
$\mu$=-1 which is invariant with the isotope substitution.
The bare phonon energy is set as $\omega_0$=1.0, and the two $e$-ph coupling
constants are fixed at a ratio of $s_1$:$s_2$=1:-1, for simplicity.
In Fig. 2, we plot the electronic dispersion relation (solid curve) along
the symmetry line $M$-$\Gamma$-$Y$ of the Brillouin zone below the Fermi
energy ($E_F$), when $s_1$=0.05 and $\lambda$=1.
The Fermi surface of this case is also shown in the inset.
By comparing with the results of Ref. 16 (dashed curve in Fig. 2), one
can see the band structure of Bi2212 has been qualitatively depicted
by our selected parameters.

\begin{figure}
\includegraphics{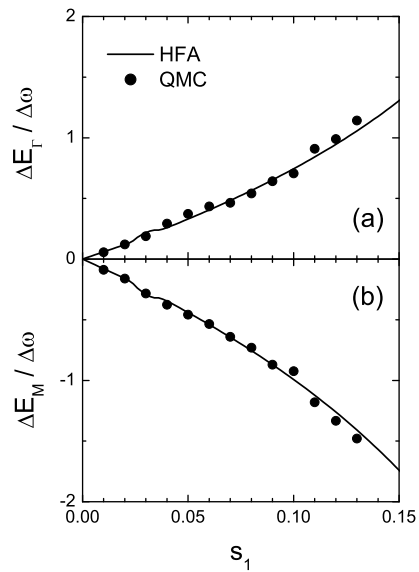}
\caption{
The positive-negative sign change effect of isotopic band shift as a function
of the $e$-ph coupling constant $s_1$, on a 4$\times$4 square lattice.
The ratio $\Delta E_{\bf k} / \Delta \omega$ at $\Gamma$ (upper panel)
and $M$ (lower panel) points are shown, where the solid curves are
calculated by HFA, and the filled circles by QMC.
}
\end{figure}

\begin{figure}
\includegraphics{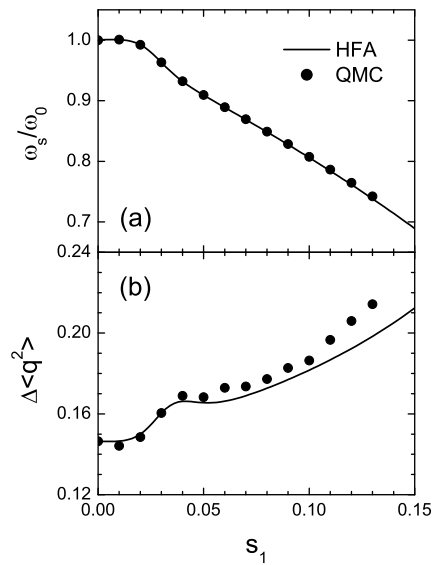}
\caption{
The changes of $\omega_s / \omega_0$ (upper panel) and phonon spatial
variation $\Delta \langle q^2 \rangle$ (lower panel) with $s_1$, on a
4$\times$4 square lattice.
The solid curves are calculated by HFA, and the filled circles by QMC.
}
\end{figure}

In Fig. 3, the HFA (solid lines) and QMC (filled circles) results of
the ratio $\Delta E_{\bf k} / \Delta \omega$ at $\Gamma$
[${\bf k}_{\Gamma}$=(0,0)] and $M$ [${\bf k}_M$=($\pi$,0)] points are
presented as functions of $s_1$, respectively.
Here we see both theories figure out an increase of
$\Delta E_{\bf k} / \Delta \omega$ with $s_1$,
which means if the $e$-ph coupling is strong enough, a small change of
phonon energy $\Delta \omega$ can be amplified to a large band shift
$\Delta E_{\bf k}$ in ARPES.

In Fig. 4, we show the origin of this anomalously large band shift is the
phonon softening driven by the aforementioned quadratic $e$-ph coupling.
In panel (a), the ratio $\omega_s / \omega_0$ is plotted as a function
of the quadratic coupling constant $s_1$.
Both HFA (solid lines) and QMC (filled circles) show that
$\omega_s / \omega_0$ declines monotonically with the increase of $s_1$.
Together with this softening of phonon energy, the phonon spatial variation
$\Delta \langle q^2 \rangle$ due to the isotope substitution increases
gradually [panel (b)].
This finally leads to a large band shift, as manifested in Eq. (12).

In Eq. (12) as well as Eq. (27), it is implied that the band shift has
a momentum dependence, which may result in an anisotropy of the isotope
effect, and a positive-negative sign change of the ratio
$\Delta E_{\bf k} / \Delta \omega$.
As an example, we show this sign reverse in Fig. 3, where panels (a) and
(b) are for $\Delta E_{\bf k} / \Delta \omega$ at the $\Gamma$ and $M$
points, respectively.
Because of the size limitation of QMC, we do not have much data for
a continuous dispersion along the $\Gamma M$ direction.
Nevertheless, in Fig. 3, both theories clearly show that the isotope
substitution yields a positive band shift at the $\Gamma$ point, and a
negative one at the $M$ point.
Thus, one can see the sign reverse of the isotope effect is a natural
consequence of the momentum dependence of the $e$-ph coupling.

In Figs. 3 and 4, although at larger $s_1$, some deviation between the
HFA and QMC arises, the overall tendencies are in a good agreement,
demonstrating that the strong off-diagonal quadratic $e$-ph coupling is
responsible for the large isotopic band shift.

Before closing this section, it is worth noting that the latest experiment
by Douglas {\it et al.}\cite{do07} has suggested that the unusual isotope
effect observed by Gweon {\it et al.}\cite{gw04} might be an alignment
error of the sample.
In contrast to the 10-40 meV band shift reported in Ref. 4, new data
show the shift is only 2$\pm$3 meV.\cite{do07}
If this is the case, it should be clear that the off-diagonal quadratic
$e$-ph coupling investigated in this work cannot be very strong for the
cuprates.

\section{Summary}

In conclusion, by using the HFA and QMC, we study the isotope effect of
Bi2212 based on an off-diagonal $e$-ph model, where the electrons and
phonons are coupled in a quadratic form.
The HFA enables us to capture the essence of present problem, and QMC
provides us the exact results with which the HFA results can be compared.
Our calculation demonstrates that the isotope-induced band shift of ARPES
can be well described by the model with an off-diagonal quadratic $e$-ph
coupling.
We ascribe the large value of $\Delta E_{\bf k} / \Delta \omega$ to the
phonon softening triggered by the quadratic $e$-ph coupling.
While the positive-negative sign change is connected to a momentum
dependence of the coupling.
Comparing the results of HFA and QMC, we find the agreement is good.

\end{document}